\documentclass[journal]{IEEEtran}
\usepackage{blindtext}
\usepackage{amsmath}
\usepackage{graphicx}
\usepackage{color}
\usepackage{gensymb}

\usepackage[english]{babel}
\usepackage[utf8]{inputenc}
\usepackage{algorithm}
\usepackage[noend]{algpseudocode}

\ifCLASSINFOpdf
\else
\fi
\begin{document}

%\title{A Unified Structure: Human Object Detection and Tracking Algorithm at the Edge of the Network}

\title{Kerman: A Hybrid Lightweight Tracking Algorithm to Enable Smart Surveillance as an Edge Service}


\author{\IEEEauthorblockN {Seyed Yahya Nikouei${^\dagger}$, Yu Chen${^\dagger}$, Sejun Song$^\xi$, Timothy R. Faughnan${^\ddagger}$} 

\IEEEauthorblockA{${^\dagger}$Dept. of Electrical and Computing Engineering, Binghamton University, SUNY, Binghamton, NY 13902, USA\\
${^\xi}$School of Computing and Engineering, University of Missouri-Kansas City, Kansas City, MO 64110 USA\\
${^\ddagger}$New York State University Police, Binghamton University, SUNY, Binghamton, NY 13902, USA\\}
E-mails: \{snikoue1, ychen, tfaughn\}@binghamton.edu, songsej@umkc.edu.

%\thanks{$^\ast$Corresponding Author: Y. Chen, Tel: (607) 777 6133, email: ychen@binghamton.edu.}
}

\maketitle

\begin{abstract}
Edge computing pushes the cloud computing boundaries beyond uncertain network resource by leveraging computational processes close to the source and target of data. Time-sensitive and data-intensive video surveillance applications benefit from on-site or near-site data mining. In recent years, many smart video surveillance approaches are proposed for object detection and tracking by using Artificial Intelligence (AI) and Machine Learning (ML) algorithms.  However, it is still hard to migrate those computing and data-intensive tasks from Cloud to Edge due to the high computational requirement.
In this paper, we envision to achieve intelligent surveillance as an edge service by proposing a hybrid lightweight tracking algorithm named Kerman (Kernelized Kalman filter). Kerman is a decision tree based hybrid Kernelized Correlation Filter (KCF) algorithm proposed for human object tracking, which is coupled with a lightweight Convolutional Neural Network (L-CNN) for high performance. The proposed Kerman algorithm has been implemented on a couple of single board computers (SBC) as edge devices and validated using real-world surveillance video streams. The experimental results are promising that the Kerman algorithm is able to track the object of interest with a decent accuracy at a resource consumption affordable by edge devices. 

\end{abstract}


% \fi
\IEEEpeerreviewmaketitle

\section{Introduction}

The unprecedented pace of urbanization \cite{chen2016dynamic} poses many opportunities and challenges. The recent concept of Smart Cities has attracted the attention of the urban planners and researchers to enhance the security and well-being of the residents. One of the most essential smart community services is the intelligent resident surveillance \cite{cenedese2014padova}. It enables a broad spectrum of promising applications, including access control in areas of interest, human identity or behavior recognition, detection of anomalous behaviors, interactive surveillance using multiple cameras and crowd flux statistics and congestion analysis and so on \cite{hu2004survey}. 

Many of these smart surveillance applications require significant computing and storage resources handling massive contextual data created by video sensors. The cloud computing paradigm provides excellent flexibility and is also scalable corresponding to the increasing number of surveillance cameras. In practice, however, there are significant hurdles for the remote cloud-based smart surveillance architecture. Many key surveillance applications such as monitoring and tracking need a real-time capability. However, processing raw video data from widely distributed video sensors such as Close-Circle Television (CCTV) cameras and mobile cameras not only incurs uncertainty in data transfer and timing but also poses significant overhead and delay to the communication networks\cite{chen2016smart}. Also, it may cause the data security and privacy issues by providing more attacking opportunities for adversaries. Therefore, current surveillance applications are for off-line forensics analysis instead of a proactive tool to deter suspicious activities before the damages are caused. 

Edge computing as a surveillance service is considered as the answer to the shortcomings\cite{ahmed2017mobile}, \cite{ali2017sedasc}, \cite{cao2017edge}. The edge computing technology migrates more computing tasks to the connected smart “things” (sensors and actuators) at the edge of the network \cite{shi2016edge}. Consequently it possesses the following advantages: \textit{real-time response}, \textit{lower network workload}, \textit{lower energy consumption}, and \textit{higher data security and privacy}.

Despite the promising Edge computing benefits, one of the critical challenges is how to efficiently process the data-intensive tasks in real-time on a rather resource hungry edge nodes. Specifically, many smart video surveillance approaches for object detection and tracking propose to use Artificial Intelligence (AI) and Machine Learning (ML) algorithms.  However, they usually have the high computational requirement. How to migrate those computing and data-intensive tasks to the edge nodes are still significant challenges.

In this paper, a novel lightweight, hybrid tracking algorithm named Kerman is proposed, which addresses the weaknesses of the well-known Kernelized Correlation Filter (KCF) \cite{henriques2015high} by leveraging some features of Kalman Filter \cite{patel2013moving} and Background Subtraction (BS) \cite{nguyen2016human}. The proposed Kerman algorithm achieves higher performance while preserves the favorite features of KCF such as its fast adaptation and tracking. An experimental study has been conducted using real-world surveillance video streams on two types of single board computers (SBC) as edge devices, Raspberry PI 3 and Tinker board. 

The rest of the paper is organized as follows: in section \ref{sec:related work} the previous attempts for object tracking at the edge of the network is discussed. %Section \ref{sec:KCF} discusses the rationale of the KCF algorithm. Sections \ref{sec:KF} and \ref{sec:BS} steps through the components to build the Kerman algorithm on top of the KCF for a better performance. 
Section \ref{sec:algorithm} presents the complete Kerman tracking algorithm. Experimental results of the tracker is presented in Section \ref{sec:experimental}. Finally, Section \ref{sec:conclusions} wraps up this paper with the conclusions.

%%%%%%%%%%%%%%%%%%%%%%%%%%%%%%%%%%%%%%%%%%%%%%%%%%%%%%%%%%%%%%%%%%%%%%%%%%%%%%%%%%%%%%%%%%%%%%%%%%%%%%%%%%%%%%%%%%%%%%%%%%%%%%%%%%%%%%%%%%%

\section{Motivation and Related Work}
\label{sec:related work}

Currently, most of the video surveillance systems function as an archive of footages and being used for the afterwards, offline forensics analysis as well as depending on human operators in process loop \cite{chamasemani2013systematic}. Because of long transmission and process time, it cannot support uninterrupted, real-time video surveillance tasks. However, thanks to the fast development in machine learning (ML) algorithms, there are very promising results presented in human-oriented surveillance area. Where human detection and tracking along with abnormal behavior detection analysis are feasible using various smart deep learning or other ML algorithms. There are various methods for automated video frames collection in a cloud and unusual events detection \cite{vishwakarma2013survey}. 
%In most surveillance systems today, processing and analyzing the collected video is pushed to cloud centers where abundant computing resources are allocated. This means all video needs to be transmitted to remote servers no matter they are processed online or offline. 
Research community has recognized that heavy communication overhead is not tolerable in many delay sensitive, mission-critical tasks \cite{shi2016edge}. Leveraging the fog computing paradigm, there are online and uninterrupted target tracking systems proposed to meet the requirements of real-time video processing and instant decision making  \cite{chen2016dynamic}, \cite{mukherjee2018survey}. %For instance, Chen et al. in \cite{chen2016dynamic} merges raw video stream generated by drones on near-sight fog computing devices, such as tablet or laptop.

While there are a lot of work conducted in image processing area such as object detection and tracking, only few literature is specifically focused on human object \cite{nikouei2018intelligent}. As the first step for any video surveillance application, object detection and coordinates allocation are essential for further object tracking tasks. CNNs have high accuracy  as well as shorter run time after training \cite{cristani2013human}, \cite{pang2012robust}. Smaller size networks are tailored for edge constrained environment to resolve the issue of limited memory  \cite{howard2017mobilenets}, \cite{iandola2016squeezenet}. For instance, a lightweight Single Shot Multi-box Detector (SSD) CNN is chose for human detection and decent performance is achieved \cite{nikouei2018lcnn}. However, even lightweight CNNs that are designed for low power devices, are not fast enough to perform real time object detection, the best ones reached about 2 FPS in experiments on Raspberry PI 3 \cite{nikouei2018lcnn}. 

This yields for a tracker that can follow a human being once it is identified and does not mix it with other moving objects in the frame. Online trackers are preferred since they conduct training online for better results. This is critical because people subject for tracking may with any size and clothes colors, offline trained systems are not suitable. Meanwhile, the tracking tasks will take place at the edge environment in which it is assumed there is not GPUs, therefore, the CPU based tracking algorithms are subject of this research.  

Analytically, a smart surveillance task can be considered as a three-layer framework: 

\begin{itemize}
\item Layer 1: the low-level conducts information extraction like feature detection and object tracking;
\item Layer 2: the intermediate-level is in charge of mode recognition like action recognition and behavior understanding; and
\item Layer 3: the high-level is focused on decision making like abnormal event detection.
\end{itemize}

\begin{figure}[t]
    \centering
        \includegraphics[width=0.425\textwidth]{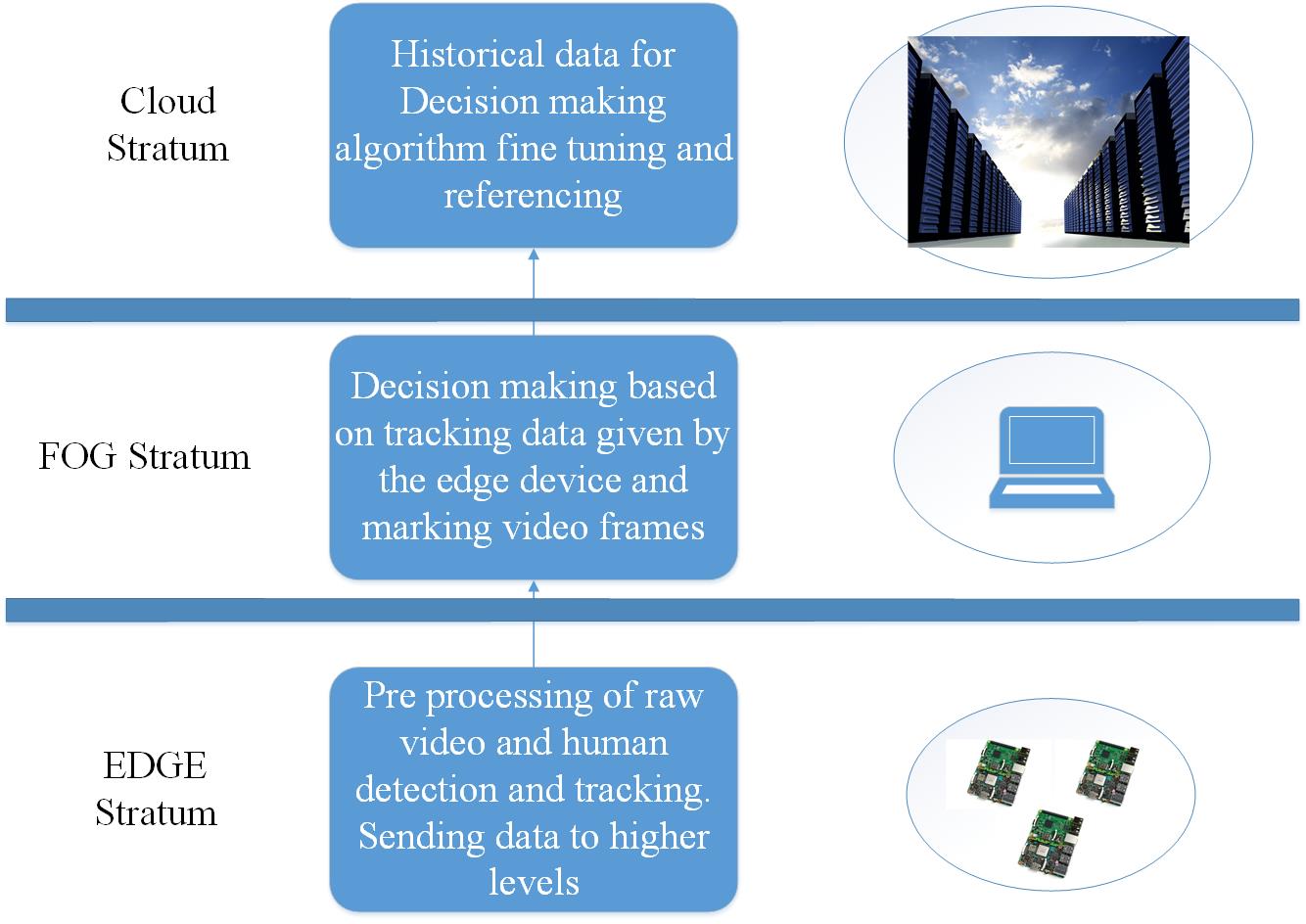}
    \caption{Human surveillance algorithm implementation at edge technology.}
    \label{fig:implementation}
    \vspace{-10pt}
\end{figure}

Functions belonging to each layer may be deployed on different positions of the hierarchical platform consisting of edge, fog, and cloud stratum \cite{mouradian2017comprehensive}. Most of the first layer functions are expected to run on the edge devices, like the surveillance cameras, leveraging light weighted algorithms. Figure \ref{fig:implementation} shows how each layer of the human surveillance system. The detection and tracking algorithms can be implemented at the edge layer with minimum latency and online training for better performance. It should be noted that in human surveillance application, there may be multiple objects being tracked in one individual frame. The detection algorithm may give the bounding box around a person but it will not guarantee the order of detection, which implies the object that was labeled in one frame may be re-labeled differently and be refereed to as another object in the following frame. 

For trackers the common challenges include partial or full object occlusions, scene illumination changes and object shapes and motion \cite{ojha2015image}. The well-known region based tracking algorithm detects a human and extracts it from the background \cite{wren1997pfinder}. Because it only detects the background and foreground based on Gaussian modeling, the region based tracking is not suitable for surveillance application at hand. Another method is Feature Based Tracking, where a classifier looks for features that are well-describing the object of interest such as lines, point that separates the object from the background. The feature based methods suffer from occlusion problem as the they need at least some sub-features to remain visible and even then the accuracy of the classification drops \cite{wu2015object}. Active Contour Based Tracking represents object's outlines as bounding contours \cite{o2010rear}.

In 2017 Need for Speed (NFS) was introduced as a dataset created with very high quality videos used for benchmarking and divides object tracking to deep trackers and correlation filter (CF) trackers \cite{galoogahi2017need}. Several best performing algorithms are used in the benchmarks. The fastest algorithm is the Multi-Dimensional Network (MDNet) that has more than 50 FPS tested using the benchmark \cite{nam2016learning}. In contrast, the CF trackers like Multiple Instance Learning (MIL) \cite{babenko2009visual} and Boosting algorithm \cite{grabner2006real} are slow. The MOSSE filter \cite{bolme2010visual} is very fast but not accurate. The KCF \cite{henriques2015high} is based on MOSSE too but it achieved better accuracy with supports from the HOG features. Because of the boundary issues in frequency domain learning \cite{galoogahi2015correlation}, some researchers use boundary learning methods to reach good performances. KCF has a much higher speed on CPU than the others without sacrificing the accuracy. CF trackers using CNN achieved a high accuracy with a low speed. 

\section{The Kerman Algorithm}
\label{sec:algorithm}

\subsection{Design Rationale}

In this effort the edge computing paradigm is leveraged for real-time human targets detecting and tracking. All the raw video streams are processed locally by edge devices instead of being sent to the remote cloud center over the communication network. While there are several methods for object tracking, only few are feasible for edge implementation. The KCF algorithm is considered as the foundation because of its fastness and light weight. It has the potential to serve the purpose of real-time surveillance at the edge. The Kernel Trick and matrix multiplications in frequency domain provide the option that enables faster computation and thus the ability to use more complex classifiers online. % once we are able to tackle its shortcomings, such as object loss in case of occlusions or getting behind when tracking a fast moving object. Specifically, multiple lightweight methods are investigated for the best positioning of bounding box that represents the object of interest while meeting the real-time requirement. 
Meanwhile, the KCF has weaknesses to be addressed. It loses the object of interest if the object moves fast, and this flaw exits no matter how well the algorithm is implemented. As the KCF algorithm considers the background in the coordinates given in each frame as well as the object. If the object of interest moves fast, soon the tracker will be mistakenly trained to focus on the background and lose the object of interest. %Human object tracking is more complex as there are many challenging scenarios, such as out of line human movement and occlusions that can interrupt the tracking . Researchers proposed to solve this problem using the depth information of the stereo cameras  \cite{hannuna2016ds}. However, nowadays it is common that there is only one single camera installed at each surveillance site such that the depth information cannot be extracted.
Additionally, the tracker tends to stop when the pedestrian walks with the usual speed but there is a sharp border line of another object or shadow, which may block the human object of interest. Therefore, another method for occlusions detection is needed. %For this purpose, the Kalman Filter (KF) \cite{patel2013moving} is considered.

Kalman Filter (KF) \cite{patel2013moving} is one of the most popular object tracking methods. If the object of interest is viewed as a system with the central point in the bounding box as its representation, its position in the next frame can be predicted. Running other code along with KCF algorithm may improve the accuracy of the tracker at the cost of slightly lower speed. KF needs to be fed with the actual position of the object of interest in each frame to create a feedback system and update its parameters. %Hence, the human detection algorithm needs execution for each frame \cite{patel2013moving}, \cite{salhi2012object}. The edge environment is constrained in terms of the number of times the human detection algorithm can be run each second. The experimental studies on Raspberry PI as edge device showed that no more than two frames can be processed per second \cite{nikouei2018intelligent}, \cite{nikouei2018lcnn}. Therefore, it is infeasible to run the KF update for every frame. 
However, the KF can be considered as a post processing for KCF algorithm, and the data from the KCF algorithm is used for KF update. If the KCF bounding box stops because of an error in tracking, the KF will follow with a delay because of its nature to remain in the situation it was before update. This delay creates a distance from the center of bounding box in the KCF and the output of the KF, which can be used as a pointer to indicate the occlusions and to prevent the KCF algorithm from mistakenly re-focusing on the background or other items but the object of interest. %The KF algorithm does not create loops and the matrices are $4 \times 4$ at most. Their size will not change based on the object's appearance. The computational complexity of KF is $O(n)$, where $n$ is the number of the objects to be tracked in each frame.  

By its nature, the move of a person is hard to be predicted, a person can make sudden changes either in terms of speed or direction. These variations create the same challenges as what occlusions introduce. On the one hand, the KCF is a very capable algorithm to follow the object with sudden changes in appearance or moving direction. On the other hand, KF lacks this ability and has a delay to follow, but this "shortcoming" is useful to stop KCF from launching wrong updates. The key here is a intelligent decision making that help the system to choose between KCF or KF correctly. In this paper, the background subtraction (BS) method based on Gaussian Mixture-based Background/Foreground Segmentation Algorithm \cite{zivkovic2006efficient} is selected to address this issue.

Basically, a bounding box is given by the KCF algorithm, the pixels from the mask (classified as foreground) near the KCF bounding box are considered as the object of interest. %It is worth noticing that the GMM model may show some single pixels as foreground, these can be treated as noises and can be prevented using Gaussian smoothing functions or other methods of noise reduction. Also, one extra filter is implemented at the output of the algorithm to discard contours smaller than four pixels. 
The background subtraction is going to be executed only one time in each frame. The algorithm complexity is $O(n)$ where $n$ is the number of modeling pixels. In the framework of our proposal, the background subtraction is not used for object tracking. Instead, it functions as an indicator telling the system whether or not the KCF or KF should be applied. The bounding boxes that are output of the background subtraction can be in any position of the frame where no human is detected. Thus, it is important to associate them to each human that is being tracked. If the center of each contour is in range of the KCF bounding box, the background subtraction verifies the human object is subject of tracking.

\begin{figure}[t]
   \centering
        \includegraphics[width=0.5\textwidth]{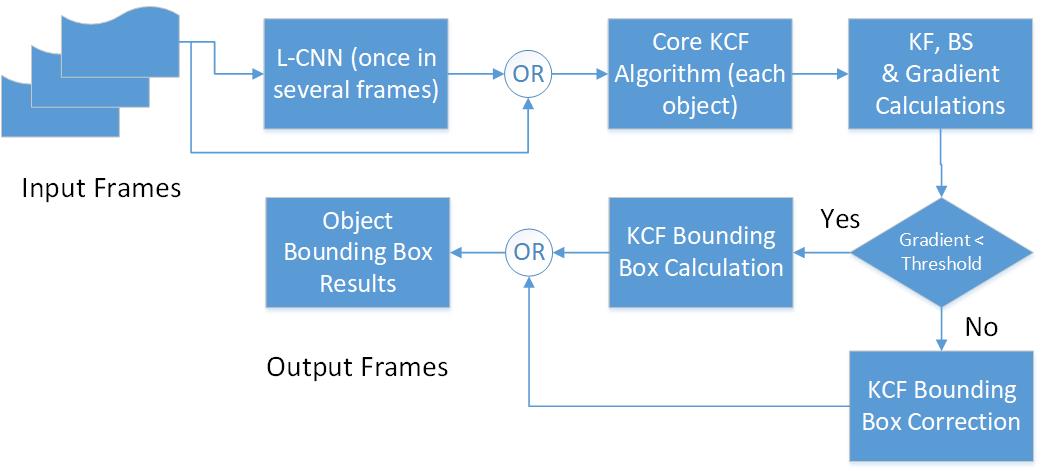}
   \caption{The overall algorithm work-flow.}
    \label{fig:work_flow}
   \vspace{-10pt}
\end{figure}

\subsection{The Kerman Algorithm: Pseudocode}

The proposed Kerman algorithm is an answer to the flaws in KCF algorithm. While each individual method has some overhead, the combination is not as heavy as simply cascading them in series. 

The Kerman algorithm is designed based on the knowledge that tracking with an online learning method is the best way to make the tracker adaptive to the object of interest and make it more immune to sudden changes to the object's appearance. In surveillance applications, human beings are walking and it is possible the pedestrian changes the directions swiftly. Consequently, the tracker may lose the target. Because it changes features that the tracker is using. 

Pseudo code of the proposed tracker is presented in Algorithm \ref{table:algorithm}. The Kerman algorithm actually makes decisions based on a majority voting mechanism, taking into account of the opinions of KCF, KF, and BS. These algorithms work together for each situation to stop KCF training, recalibrate the bounding box or continue normal path. In order to use center coordinates that are obtained by the KCF, by default the tracker is set as KCF. Next, the center of the bounding box given by the KCF is set as coordinate $(0,0)$ for each specific object. Knowing the centers of KF and BS algorithms for the same object can give two gradient for the corresponding object and a threshold between gradients gives the flag to recalibrate KCF bounding box. In should be noted that in the Kerman algorithm, a class $tracking \_ object$ is introduced, objects from this class are created in a multi-thread manner in order to utilize the multi-core processor of the edge device. The number of threads depends on the number of cores the edge device has. %, in our case four threads is chosen. 

\begin{algorithm}
\caption{Decision\_Tree\_Tracker}
\label{table:algorithm}
\begin{algorithmic}[1]
\Procedure{$Decision\_Tree$}{new\_coord}
\State $feature \gets bbx\_area.HOG()$
\If {flag}
	\State $KCF.training(features)$ 
\EndIf
\If {cls\_pxl $\gets$ 1.5$\times$bbx.area()}
	\State $bbx \gets KCF.classify$ \Comment{pixles in $\times$1.5 the area of given bbx (comes from previous frame) using the trained model}
\Else
	\State $bbx \gets KCF.classify$ \Comment{the area of the bbx}
\EndIf
\State $cnt\_KCF \gets KCF.center(bbx)$
\If {flag}
	\State $cnt\_KF.update(bbx)$
\Else
	\State $cnt\_KF.update()$
\EndIf
\State $KF\_grad \gets gradient(cnt\_KCF,cnt\_KF)$
\For {each cnt\_cont}
	\If {cnt\_cont in bbx}
		\State $bs\_grad \gets gradient(cnt\_KCF,cnt\_cont)$
	\EndIf
\EndFor
\If {bs\_grad - KF\_grad < trhd \textbf{and} bs\_grad found}
	\State $flag \gets TRUE$
    \State $bbx \gets bbx(cnt_KF,cnt_cont)$
    \Comment {the bounding box is changed to have center same as middle of KF and BS centers}
\ElsIf {bs\_grad \textbf{not} found}
	\State $bbx \gets bbx(cnt_KF)$
    \Comment {the bounding box is changed to have center same as middle of KF}
\Else
    \State $flag \gets FALSE$
    \Comment {bounding box is given by the KCF}
\EndIf
\Return {$bbx$}
\EndProcedure

\Procedure{Tracker}{camera1.stream} 
\State $frame \gets camera1.stream$ \Comment{OpenCV library}
\While{frame not empty}\Comment{camera on}
	\State $frame\_400 \gets frame.resize(400\times400)$
	\If {$(framecount> human\_chk\_thld)$}
		\State $fnd\_obj \gets feed(L-CNN(frame\_400))$ \Comment{human detection CNN based feed-forward}
		\For{each fnd\_obj}
        	\For{each trk\_obj}
            	\State$checking weather...$ 
                \State $coordinates are new$    
            \EndFor
            \State $Decision\_Tree.trk\_obj\_Q  \gets new\_coord$ \Comment{add new object to the queue}
		\EndFor
	\EndIf
    \State $cont \gets bg\_sub(frame\_400)$
    \For {each Decision\_Tree.trk\_obj\_Q}
    	\State $trk\_obj.updates(frame\_400,cont)$
    \EndFor
    \State $show.frame()$
\EndWhile\label{euclidendwhile}
\EndProcedure
\end{algorithmic}
\end{algorithm}

The overall Kerman algorithm work-flow is shown in Fig. \ref{fig:work_flow}, where from the input frame the human object bounding boxes are given.

\section{Experimental Results}
\label{sec:experimental}

The proposed Kerman algorithm has been implemented on two types of Single Board Computers (SBC) for test. One is the Raspberry PI 3 with 1 GB of RAM and ARMv7 1.2 GHz processor, the other device is a Tinker Board with 1.8 GHz ARM-based RK3288 SoC and 2 GB LPDDR3 dual-channel. These SBCs are selected as the edge devices because of the low price (\textless\$80 each), but at the same time they are capable of running UNIX or Android based operating systems. Such that they support high level programming like Python and various I/O connections. Actually it is a trend that more and more powerful small devices like SBC have shown faster and reliable performances at the edge of the network.

\subsection{High Level Overview}

\begin{figure}[t]
    \centering
        \includegraphics[width=0.425\textwidth]{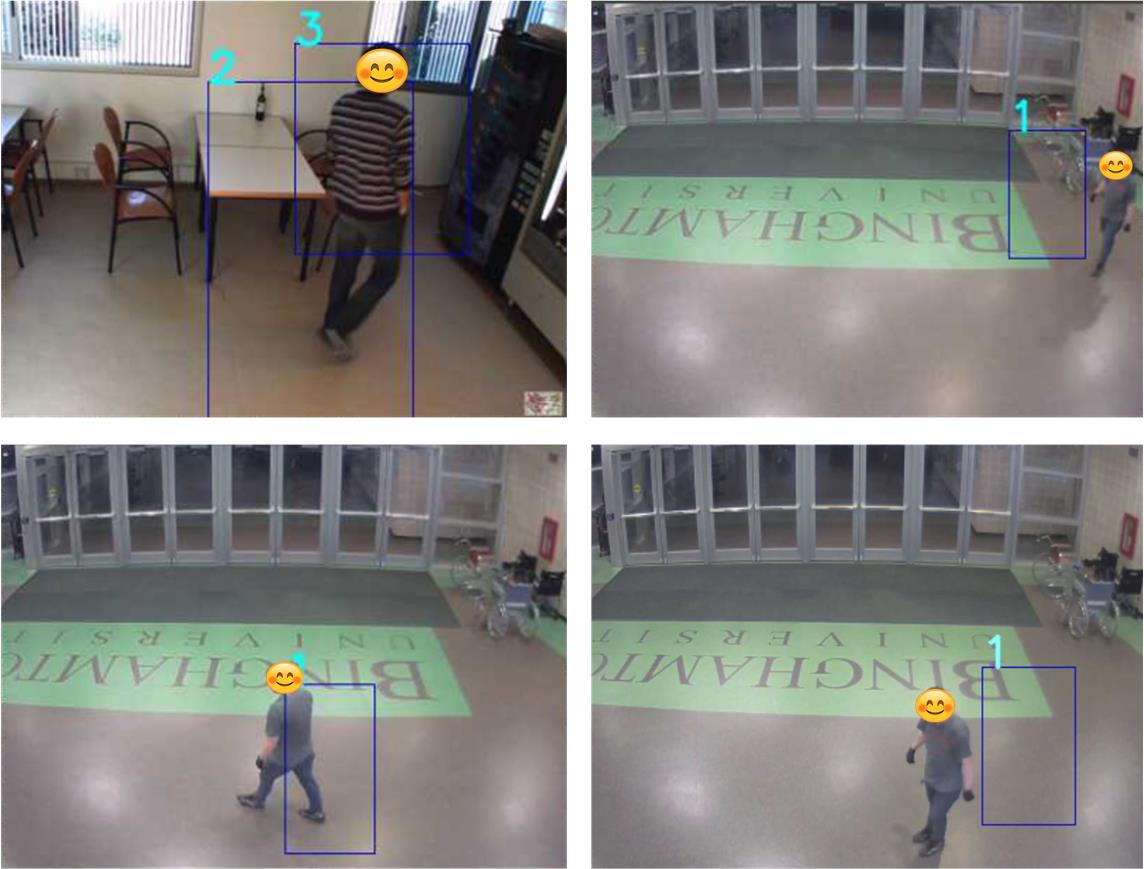}
    \caption{KCF algorithm lag when the object of interest moves fast.}
    \label{fig:object lose}
    \vspace{-10pt}
\end{figure}

\begin{figure}[t]
    \centering
        \includegraphics[width=0.4\textwidth]{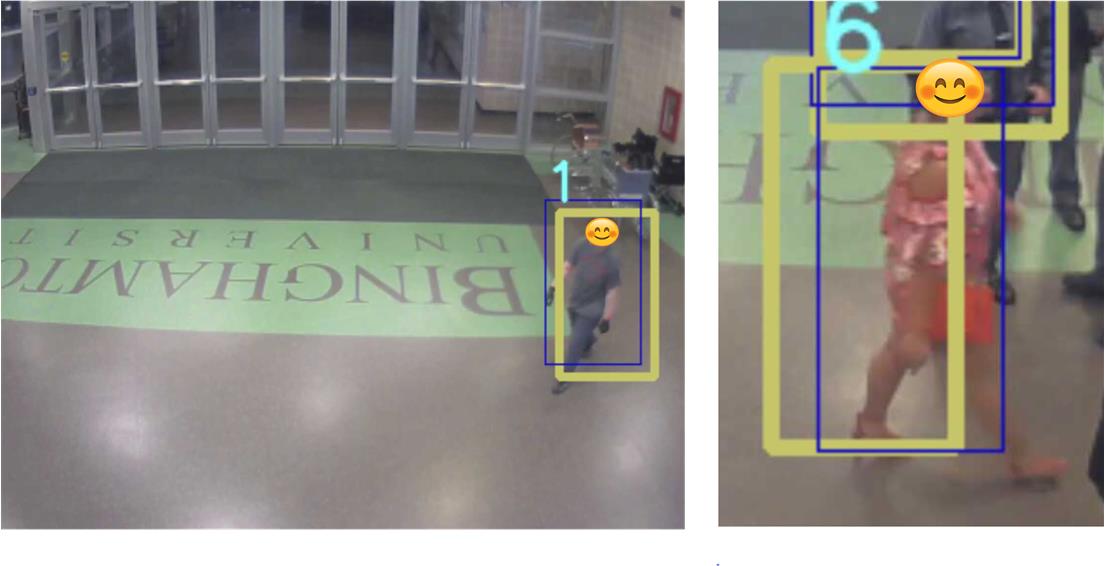}
    \caption{Kalman Filter will have a lag from KCF because KCF is used as real measurement.}
    \label{fig:KF_lag}
    \vspace{-10pt}
\end{figure}

In the tracking process, the objects that are already being tracked should not be labeled again. Therefore the center of a new detection should be out of a circle with $2/3$ diameter of the objects that are already in the queue for track. %Thus the idea of getting behind the object makes the whole algorithm fail and that is why it was one of the main focuses in here. 
Meanwhile, the non-trivial lags between the boundary box and the actual position of the object will fail the tracking algorithm. Figure \ref{fig:object lose} shows example cases in which the object was lost when using the KCF algorithm only. In the upper left image the tracker lag leads to re-detect the human and label him as a new object. This comparison is based on the KCF because to the best of the authors knowledge, it is the fastest among today's online tracking methods, which are not based on CNN. The human object detection method applied here is the L-CNN that is tailored for edge environment \cite{nikouei2018lcnn} to detect humans and pass them to the tracker. As shown in Fig. \ref{fig:KF_lag}, the KF has a lag in comparison to the KCF algorithm when the object moves fast or changes its direction abruptly. 

The proposed Kerman algorithm integrates three fast tracking algorithms, KCF, KF, and BS, to achieve a higher accuracy than each individual one can do separately. Kerman makes better decisions when the human object moves faster than the KCF can follow or there is a occlusions between object of interest and another object in the frame. Although it is true that the tracker is able to find the object of interest again using the object detection algorithm even if it lost the object under tracking, the performance is impacted significantly and the interrupted tracking will further slow down the next steps in surveillance, \textit{e.g.} anomalous behavior detection. 

\begin{figure}[t]
    \centering
        \includegraphics[width=0.45\textwidth]{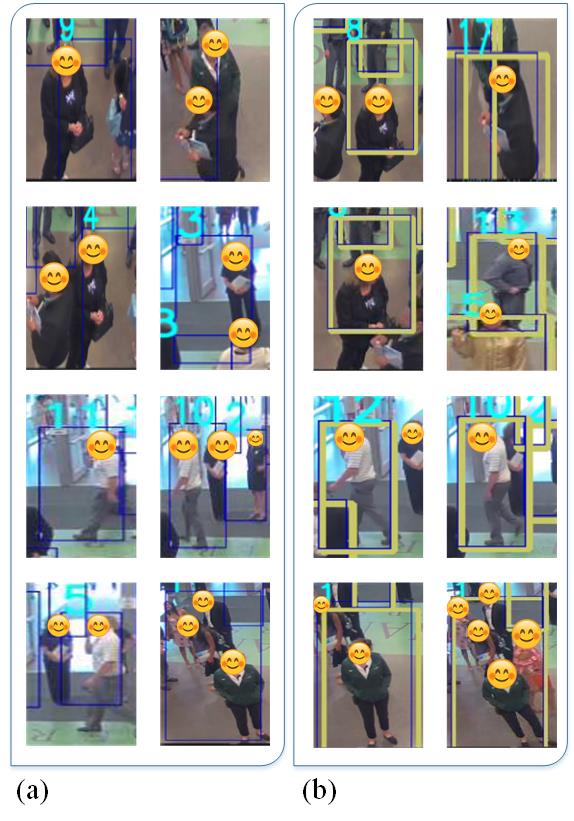}
    \caption{Performance comparison between the Kerman and KCF algorithms in case when objects move fast and occlusions exist. (a) KCF algorithm (b) KERMAN algorithm.}
    \label{fig:better_spots}
    \vspace{-10pt}
\end{figure}

%The proposed Kerman algorithm tracking decision based tree explained in section \ref{sec:algorithm} tries to 
Figure \ref{fig:better_spots} shows the results of a real life surveillance footage processing, which compares the Kerman algorithm with the KCF algorithm. Looking more closely to this figure, the part (a) shows results of instances from KCF based algorithm. Where the object is lost or the bounding box around the object has a lag or contains a huge space. In contrast, Fig. \ref{fig:better_spots}(b) shows the results of Kerman algorithm on the same video stream with same instances, but the bounding box is better fit neither the object is lost. 

\subsection{Performance Analysis}

\begin{figure}[t]
    \centering
        \includegraphics[width=0.50\textwidth]{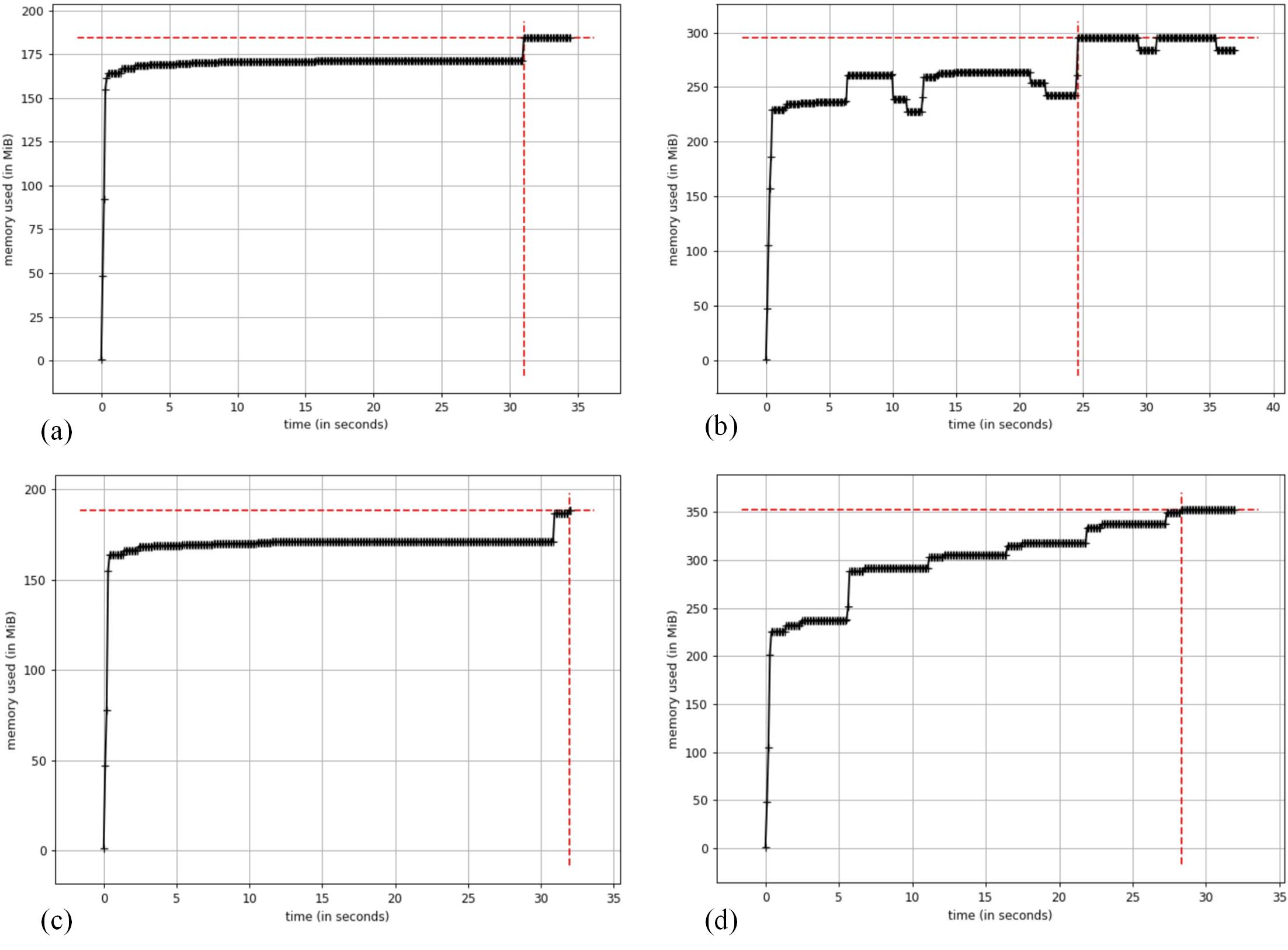}
    \caption{Memory needed in MB to run tracking and detection algorithms in 30 seconds run-time: (a) Kerman algorithm with 0-2 human objects in the frame (b) Kerman algorithm with 6-10 objects (c) KCF with 0-2 objects (d) KCF with 6-10 objects.}
    \label{fig:memory} 
    \vspace{-10pt}
\end{figure}

The performance of the Kerman algorithm has been evaluated experimentally in terms of memory consumption, CPU utility, and the video processing speed (FPS). Figure \ref{fig:memory} compares the memory consumption of the Kerman algorithm with the memory used by the KCF algorithm. The memory is read in a 30 seconds of run time of two scenarios. In one scenario, there are only one or two human objects in the frame and they are positioned far away from each other. In the other scenario, the frame is more crowded with human objects and at most 10 pedestrians are in the frame in the same time. The memory utility shown in Fig. \ref{fig:memory} includes both the tracking and detection algorithms. The human detection algorithm is the L-CNN that runs in every five frames for a video input rate of 10 FPS. The L-CNN detector runs two times per second. Considering the velocity of pedestrians, it is sufficient. In case there are up to 10 human objects in the frame the algorithm needs up to 350 MB of memory space, which is available even in a memory limited device like Raspberry PI. The experimental results also show that the memory consumption is not sensitive to the number of objects in the frame. The difference between having many objects and fewer objects is not significant, which verified the Kerman algorithm is scalable in terms of memory utility. 

The CPU usage is also a critical metrics for the detection and tracking algorithms as a whole system designed for human surveillance automation. The percentage of CPU usage is read on Raspberry PI 3 and Tinker Board for 30 seconds of runtime and averaged. Same scenarios as used for memory consumption test are applied to evaluate assess the CPU usage, but divided into two scenarios. One scenario is with at most 2 human objects in a frame and another is with 6-10 human objects. Both cases are managed using $80-89\%$ CPU.
%That is why averaging can give better intuition of the whole algorithm. 

%%%%%%%%%%%%%%%%%%%%%%%%%%%%%%%%%%%%%%%%%%%%%%%%%%%%%%%%%%%%%%%%%%%%%%%%%%%%%%%%%%%%%%%%

\section{Conclusions}
\label{sec:conclusions}

In this paper the Kerman algorithm is introduced that integrates three well-known lightweight tracking algorithms, KCF, KF and BS, to enable the smart surveillance as an edge service. The Kerman algorithm calculates the gradient of the KF and BS algorithms based on the KCF algorithm for each object of interest and recalibrates the bounding box given by the KCF. On the selected edge devices, the Kerman algorithm achieved decent performance in processing the real-world security surveillance videos. The experimental results verified that the Kerman algorithm has solved the flaws associated with the KCF algorithm at a tolerable trade-off in processing time. It meets the design goals and is a feasible solution at the edge. 

\ifCLASSOPTIONcaptionsoff
  \newpage
\fi



%\bibliographystyle{IEEEtran}

\bibliographystyle{IEEEtranS}

\bibliography{./tracker}

%\begin{biography}[{\includegraphics[width=1in,height=1.25in,clip,keepaspectratio]{mshell}}]{Michael Shell}

%\begin{IEEEbiography}%[{\includegraphics[width=1in,height=1.25in,clip,keepaspectratio]{picture}}]{John Doe}
%\blindtext
%\end{IEEEbiography}


%\vfill


\end{document}